\documentclass[aps,prl,showpacs,twocolumn,nofootinbib,amsmath,10pt,letterpaper]{revtex4-1}
\usepackage[normalem]{ulem}
\usepackage{bm}
\usepackage{epsfig}

\usepackage[latin1]{inputenc}
\usepackage{amsmath}
\usepackage{amsfonts}
\usepackage{amssymb}
\usepackage{graphicx}
\usepackage{setspace}
\usepackage[usenames]{color}
\usepackage{epstopdf}

\newcommand{\be}{\begin{equation}}
\newcommand{\ee}{\end{equation}}

\def \be{\begin{equation}}
\def \ee{\end{equation}}
\def \ba{\begin{array}}
\def \ea{\end{array}}
\def \bea{\begin{eqnarray}}
\def \eea{\end{eqnarray}}
\def \nn{\nonumber}
\def \yd{^\dagger}
\def \av#1{{\langle#1\rangle}}
\def \ket#1{{\,|\,#1\,\rangle\,}}
\def \bra#1{{\,\langle\,#1\,|\,}}

\def \expv#1{{\,\langle\,#1\,\rangle\,}}
\def \half{{\frac{1}{2}}}

\begin{document}

\title{Dissipative Preparation of Spin Squeezed Atomic Ensembles in a Steady State}

\author{Emanuele G. Dalla Torre$^{1}$}
\thanks{E.G. Dalla Torre and J. Otterbach equally contributed to this work}

\author{Johannes Otterbach$^{1}$}
\thanks{E.G. Dalla Torre and J. Otterbach equally contributed to this work}

\author{Eugene Demler$^{1}$}
\author{Vladan Vuletic$^{2}$}
\author{Mikhail D. Lukin$^{1}$}

\affiliation{$^{1}$Physics Department, Harvard University, Cambridge 02138, MA, USA, $^{2}$Physics Department, Massachussetts Institute of Technology, Cambridge 02139, MA, USA}

\date{\today}

\begin{abstract}
We present and analyze a new approach for the generation of atomic spin squeezed states. Our method involves the collective coupling of an atomic ensemble to a decaying mode of an open optical cavity. We demonstrate the existence of a collective atomic dark-state, decoupled from the radiation field. By explicitly constructing this state we find that it can feature spin squeezing bounded only by the Heisenberg limit. We show that such dark states can be deterministically prepared via dissipative means, thus turning dissipation into a resource for entanglement. The scaling of the phase sensitivity taking realistic imperfections into account is discussed.
\end{abstract}

\maketitle


The realization of spin squeezed states \cite{Kitagawa-1993} of atomic ensembles is an important subject in quantum science. Such states  play a central role in studies of many-body entanglement \cite{Lewenstein05,Briegel09,Guhne09}. In addition, they may lead to the practical improvements of state-of-the-art atomic clocks and frequency standards \cite{Preskill-2000,Hammerer-2010,Ma-review}. In a spin-squeezed-state the atoms are entangled in such a way that the fluctuations of their total spin are smaller than the sum of the fluctuations of the individual atoms. As a rule, such entangled states are extremely fragile and spin squeezing is destroyed due to  dissipation or decoherence.

In this Letter we propose a new approach for the realization of spin squeezing using recently developed ideas on quantum-bath engineering \cite{Cirac-QC,Zoller-purestate,Cirac-ent,Kastroyano,Watanabe} and show how to realize squeezing in the steady state of a dissipative atom-cavity system. In our scheme, the steady state is unique and is reached by the system starting from any initial state, without the need to adiabatically follow a particular path in the parameter space. Because the spin squeezing is achieved by optical pumping, our approach can be considerably more robust against noise as compared to existing preparation methods \cite{Kuzmich98,AndrePRL,Wasilewski-2010,AndrePRA,Polzik,Vuletic,VuleticPRA,Thompson,Mitchell} that produce short-lived spin squeezed states, limited by decoherence processes. Moreover, using properly selected atomic transitions allows to continously pump the atoms into the desired states and consequently avoid population losses due to scattering into states not not participating in the squeezing. A similar approach to continously entangle two remote atomic ensemble using free space scattering has been experimentally demonstrated in Ref.~\cite{Krauter-2011}.

\begin{figure}[t]
\centering
(a) \includegraphics[scale=0.25]{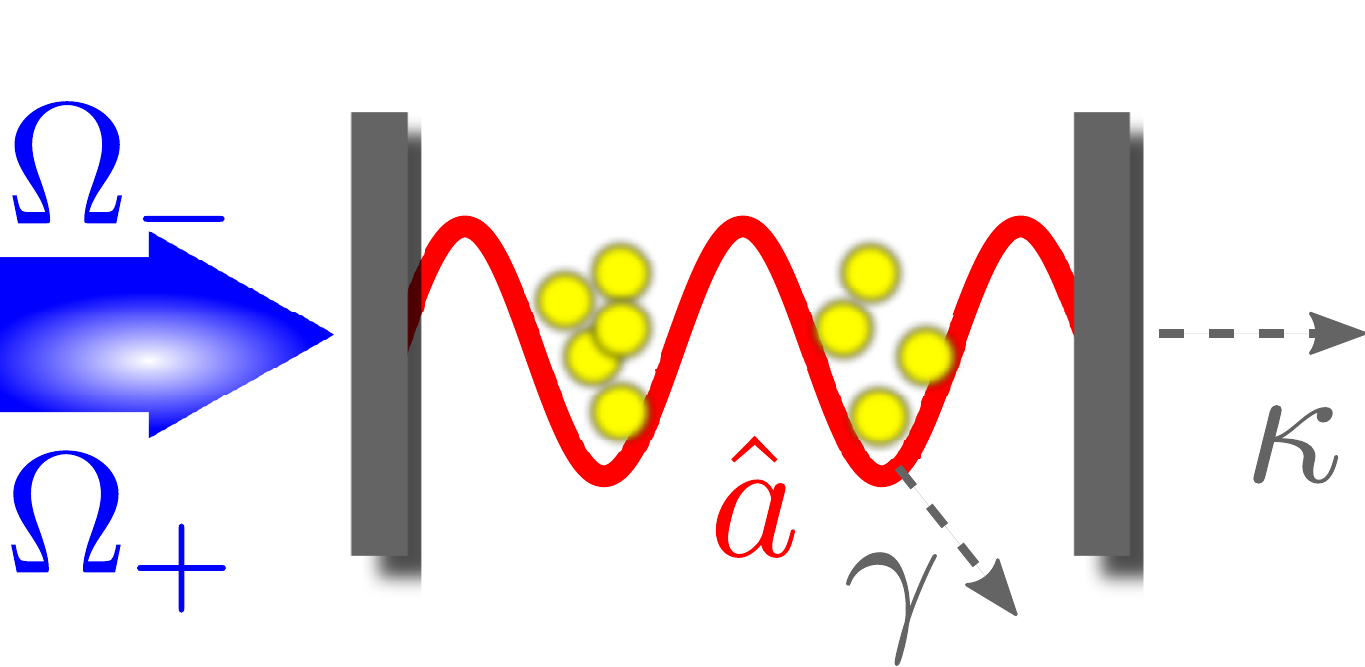}
\hspace*{.1cm}
(b) \includegraphics[scale=0.4]{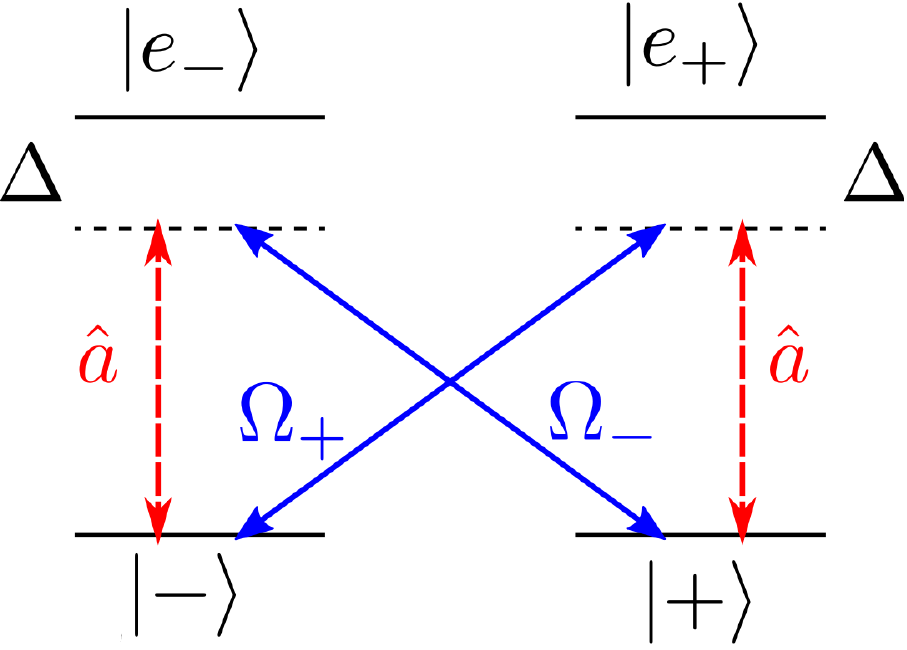}
\caption{(a) Schematic experimental setup of atoms with a spontaneous emission rate $\gamma$ in a cavity with line-width $\kappa$ supporting a single cavity mode $a$. (b) Effective linkage pattern consisting of two degenerate ground states ($\ket{\pm}$) encoding the effective atomic spin and two excited state $\ket{e_\pm}$. The system is driven via classical control fields $\Omega_\pm$ with large detuning $\Delta$ and a single global cavity mode $a$ coupling homogeneously to all atoms. {The orientation of the arrows does not necessarily correspond to the actual polarization of the fields in a specific physical realization. We give an example of such an implementation in the Supplementary Material.}
}
\label{fig:scheme}
\end{figure}

The central idea of our work can be understood by considering an ensemble of $N$ atoms interacting with a single radiation mode of an open optical cavity (Fig. 1a) and externally driven by a pair of coherent laser fields. The cavity mode and laser fields are tuned to excite a pair of two-photon Raman transitions, each involving one laser field with Rabi frequency $\Omega_\pm$ and a single cavity photon $a$ ($a^\dagger$) (Fig. 1b).  Assuming that the coupling of atoms to the cavity mode is uniform, the unitary time-evolution of such cavity-atom system is described by the effective Hamiltonian
\begin{align}
H_{\rm int} = \frac{g}{\Delta}a\yd\left(\Omega_+ S^+ + \Omega_- S^-\right)  + {\rm H.c.}\;,\label{eq:Hint}
\end{align}
where  $S^{+}$ ($S^-$) is the spin raising (lowering) operator of the collective spin, $g$ is the single atom-field coupling strength and $\Delta$ is the single-photon detuning from the excited state.
The coupling $g\Omega_\pm/\Delta$ is therefore an effective Rabi frequency corresponding to off-resonant Raman scattering (cf. Fig. \ref{fig:scheme}).
Direct examination of the Hamiltonian (\ref{eq:Hint}) shows that it has a dark-state $\ket{D}=\ket{\psi_{\rm spin}}\ket{0_{\rm cav}}$, where $\ket{0_{\rm cav}}$ is the cavity vacuum and $\ket{\psi_{\rm spin}}$ satisfies
\be
\left(\Omega_+S^+ + \Omega_-S^-\right) \ket{\psi_{\rm spin}} = 0\;.
\label{eq:dark}
\ee
The interesting features of this state are (i) that it is a pure state not containing any photons and consequently is not affected by the cavity decay, i.e. $\ket{\psi_{\rm spin}}$ is a \textit{dark-state} of the cavity decay, and (ii) it describes a highly correlated atomic state containing a high degree of spin squeezing. The latter can be seen by computing the ratio between the spin fluctuations in the $x$ and $y$ direction: $\expv{S_x^2}/\expv{S_y^2} = \expv{(S^+ + S^-)^2}/\expv{(S^+ - S^-)^2} = (\Omega_- - \Omega_+)^2/(\Omega_- + \Omega_+)^2$. For any $\Omega_+\neq\Omega_-$ this ratio is smaller than one, indicating that the dark state is spin squeezed.

\begin{figure}[t]
\centering
\includegraphics[scale=.3]{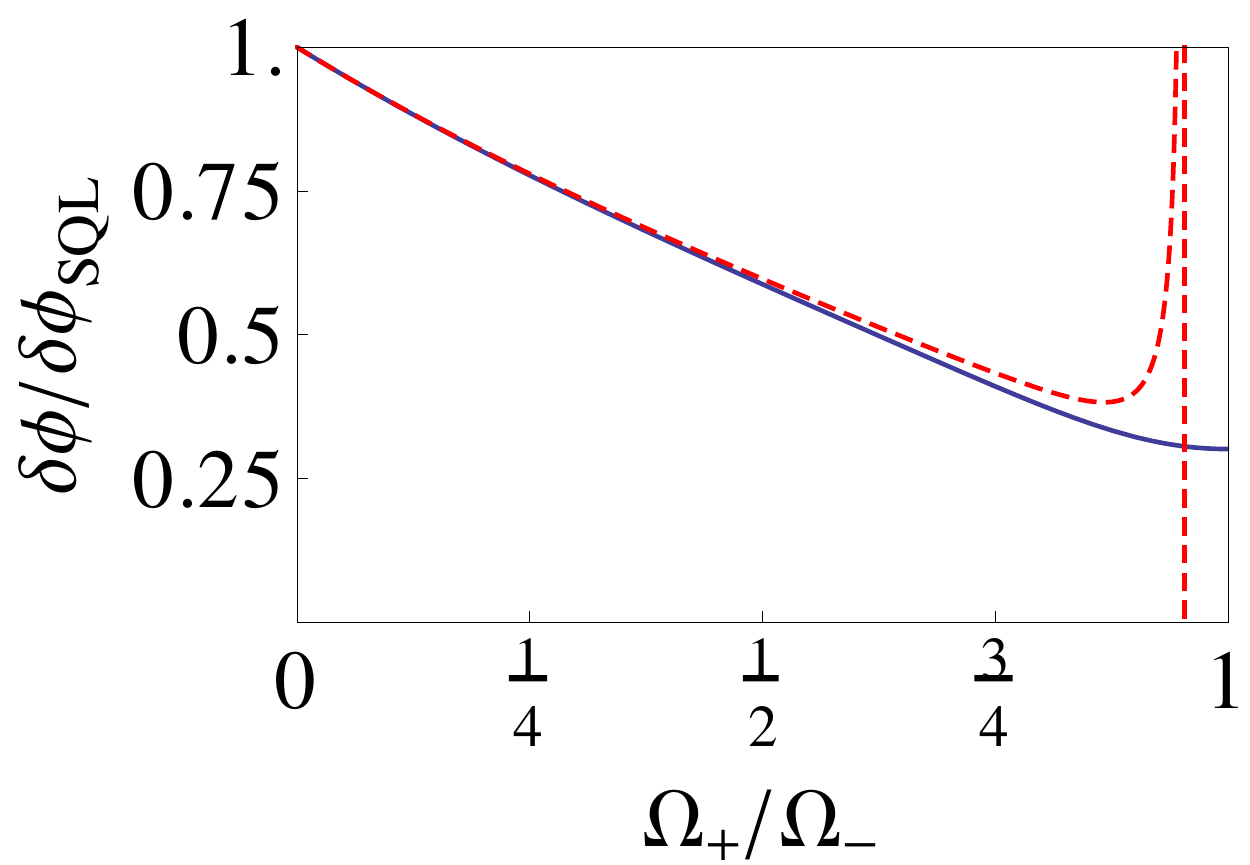}
\includegraphics[scale=.3]{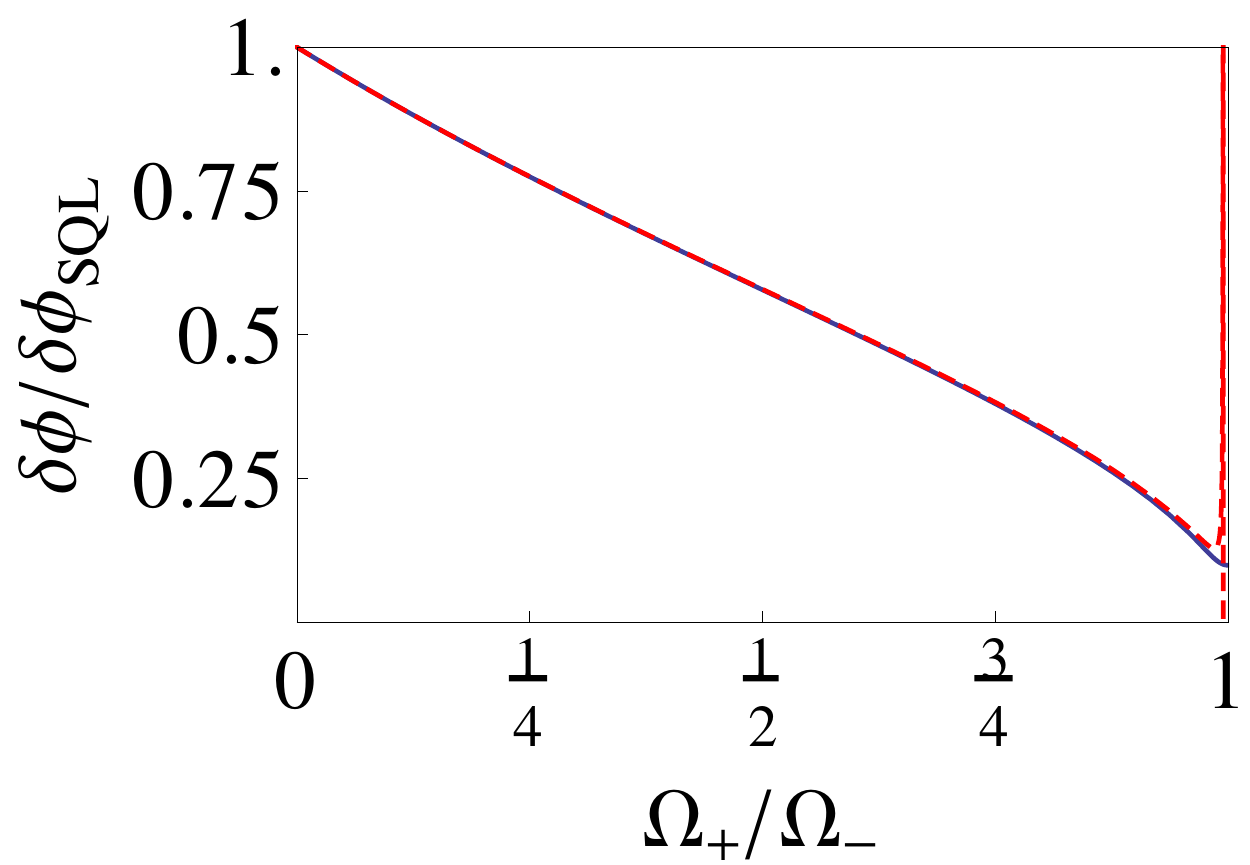}
\caption{Phase sensitivity of the dark-state $\ket{\psi_{\rm spin}}$: exact numerical solution (solid curve) and mean-field approximation (dashed curve) for $N=10$ (left) and $N=100$ (right) atoms. The optimal squeezing is obtained for $\Omega_+\to\Omega_-$ and approaches the Heisenberg limit $\delta\phi=1/\sqrt{N(N/2+1)}\sim 1/N$.}
\label{fig:singlemode}
\end{figure}

To quantify the degree of useful spin squeezing,  we use the phase sensitivity introduced by Wineland \textit{et al.} \cite{Wineland,Ma-review}
\be
\delta \phi = \frac{\sqrt{\av{S_x^2}}}{|\av{S_z}|}\;.
 \label{eq:sens}
\ee
We first focus on symmetric states with the total spin $S=N/2$, i.e. an initially spin-polarized ensemble, and expand the dark-state into eigenstates of $S_z$ as $\ket{\psi_{\rm spin}}=\sum_m c_m \ket{S=N/2,S_z=-N/2+m}$. Substituting this into (\ref{eq:dark}) results in a recursion relation between $c_{m}$ and $c_{m+2}$, in close analogy with squeezed light \cite{Yuen75}
\be
c_{m=2n} =\left(\frac{\Omega_+}{\Omega_-}\right)^n \left(\ba{c} N/2\\ n\ea\right) \left(\ba{c} N\\ 2n\ea\right) ^{-1/2} c_0.
\label{eq:exact}
\ee
Here $(:)$ are the binomial coefficients and $c_0$ is determined by the normalization condition $\sum_m |c_m|^2=1$. The phase sensitivity $\delta\phi$ of $\ket{\psi_{\rm spin}}$ is shown by the blue solid curve in Fig. \ref{fig:singlemode} as function of the ratio $\Omega_+/\Omega_-$ between the control fields. For $\Omega_+/\Omega_-=0$, the dark-state is fully polarized along the $z$-axis, giving a phase sensitivity of $1/\sqrt{N}$ corresponding to the \textit{Standard Quantum Limit} (SQL). In the opposite limit of $\Omega_+ \to \Omega_-$, the phase sensitivity approaches the \textit{Heisenberg limit} \cite{Ou-1997,AndrePRA} $\delta\phi =1/\sqrt{N(N/2+1)}$, indicating that the dark state corresponds to an almost maximally squeezed atomic state. A related method to generate spin squeezing in the weak coupling regime, i.e. $\Omega_+/\Omega_-\ll1$, has been recently proposed in Ref.~\cite{zheng12}.

We now turn to the preparation of the dark-state (\ref{eq:Hint}) and study the corresponding quantum dynamics, focusing first on the effect of the cavity decay. To this end, consider an ensemble of $N$ four-state atoms, consisting of two (meta-)stable states $\ket{+}$, $\ket{-}$ serving as our effective spin-states and two excited states $\ket{e_\pm}$. The atoms are coupled to a single mode of an open cavity (cf. Fig.\ref{fig:scheme}) with resonance frequency $\omega_a$, volume $V$ and photon escape rate $\kappa$ driving the transitions $\ket{\pm}\rightarrow\ket{e_\pm}$ and strong control fields $\Omega_\pm$ with frequency $\omega_c$ driving the transitions $\ket{\mp}\rightarrow\ket{e_\pm}$, respectively. The atom-cavity field coupling strength is given by $g=\wp/\hbar\sqrt{\hbar\omega_a/2\varepsilon_0 V}$, where $\wp$ is the dipole matrix element of the transition. We assume the control fields and the cavity mode to be far detuned from the resonant transition frequency $\omega_\text{res}=\omega_{\pm}-\omega_{e_\pm}$ between ground state and excited state manifold.
This allows us to adiabatically eliminate the excited states and we obtain the effective Hamiltonian (\ref{eq:Hint}) with $S^+ = \sum_j^N \sigma^+ = \sum_j^N \ket{+}_{jj}\bra{-}$, $S^- = \left(S^+\right)\yd$ and $|\Delta|=|\omega_\text{res}-\omega_{a,c}|$. For simplicity of the calculation we assumed here that we can realize two independent $\Lambda$-schemes according to Fig. (\ref{fig:scheme}) and neglect
any possible hyper-fine structure in the excited states. While this approximation is only true if the detunings from the respective excited state is smaller than the hyperfine-splitting in the excited state manifold, Hamiltonian (\ref{eq:Hint}) is valid for more general conditions, when Raman processes mediated by multiple virtual states, can be added up \cite{Harris1997,Harris1999}. A more detailed derivation of the Hamiltonian, also including different detunings, can be found in the supplementary material \cite{Supplementary}. Assuming $\kappa\gg|\Delta|$, we can neglect the occupation of the cavity mode and using standard techniques \cite{Gardiner-2010} we obtain the master equation in Lindblad form after tracing over the cavity modes
\begin{align}
 \frac{\partial}{\partial t}\rho=&-\gamma_{\rm cav}\left[\left\lbrace\mathcal{I}^\dagger\mathcal{I},\rho\right\rbrace_+-2\mathcal{I}\rho\mathcal{I}^\dagger\right].
 \label{eq:master2}
\end{align}
Here $\{\cdot ,\cdot\}_+$ denotes the anti-commutator, $\gamma_{\rm cav} = g^2\Omega^2/\Delta^2\kappa$ with $\Omega^2=\Omega_+^2+\Omega_-^2$, $\mathcal{I}=\sin\theta S_++\cos\theta S_-$ is the Lindblad-operator accompanying the emission of a cavity photon, and we defined $\tan\theta=\Omega_-/\Omega_+$.

To compute the phase sensitivity $\delta\phi$, we use (\ref{eq:master2}) to determine the time evolution of the expectation values of the spin operators $S_z$ and $S_x^2$
\bea
\frac{d \av{S_z}}{dt} &&= -\gamma_{\rm cav} \left[\cos^2(\theta)\av{S^+S^-} - \sin^2(\theta)\av{S^- S^+}\right]\label{eq:cavity_SZ}\\
\frac{d\av{S_x^2}}{dt} &&= -\gamma_{\rm cav} (\sin(\theta)-\cos(\theta))\label{eq:cavity_SX}\\&&\av{\left(\sin(\theta)S^- + \cos(\theta)S^+\right)(S_xS_z+S_zS_x) + {\rm H.c.}}\nn
\eea
Note that the dark state $\ket{\psi_{\rm spin}}$, defined by (\ref{eq:dark}), is indeed a steady state of (\ref{eq:cavity_SZ}-\ref{eq:cavity_SX}). Equations (\ref{eq:cavity_SZ}-\ref{eq:cavity_SX}) involve higher moments of the spin, whose time evolution should be computed independently. To avoid this complication and obtain an analytic solution, we linearize the equations of motion around $\langle S_z\rangle\approx-N/2$. Defining the small fluctuations as $\delta S_z = S_z+N/2$ and using the approximation $\langle S_x^2+S_y^2\rangle=\langle\mathbf{S}^2\rangle-\langle S_z^2\rangle \approx N\langle \delta S_z\rangle +N/2$, we find
\bea
\frac{d \av{\delta S_z}}{dt} &=& - \gamma_{\rm cav}N\left[\cos(2\theta)\av{\delta S_z} - \sin^2(\theta)\right],\label{eq:cavity2_SZ}\\
\frac{d \av{S_x^2}}{dt} &=& -\gamma_{\rm cav}N\left[\cos(2\theta)\av{S_x^2} + (1-\sin(2\theta))\frac{1}4\right].\label{eq:cavity2_SX}
\eea
The solutions to these equations decay exponentially in time with an effective rate $\gamma_{\rm eff} = \gamma_{\rm cav}\cos(2\theta) N$. From eqs. (\ref{eq:cavity2_SZ}-\ref{eq:cavity2_SX}) we obtain the phase sensitivity in the steady state as
\be
\delta\phi^2 = \frac{N(1-\sin(2\theta))\cos(2\theta)}{\left(N\cos(2\theta)-2\sin^2(\theta)\right)^2}.
\label{eq:singlemode}
\ee
In Fig.\ref{fig:singlemode} we compare this approximate expression (dashed curve) with the exact result obtained from Eq.~(\ref{eq:exact}) (solid curve). The two curves significantly deviate only for $\Omega_+/\Omega_- \approx 1 - {\mathcal O}(1/N)$, or equivalently $\theta \approx \pi/4 - {\mathcal O}(1/N)$. In this regime the present linearized approximation fails because $\av{\delta S_z} = \sin^2(\theta)/\cos(2\theta) \approx {\mathcal O}(N)$. Note that, as approaching the $\Omega_+=\Omega_-$ limit, the steady-state spin squeezing reaches its maximal Heisenberg-limited value. But, at the same time, the effective dark-state pumping rate $\gamma_{\rm eff}$ tends to zero, making the spin squeezing process extremely slow. This renders competing processes, as e.g. spontaneous Raman scattering of individual atoms, very important as will be discussed below.

\begin{figure}[t]
\centering
\includegraphics[scale=0.9]{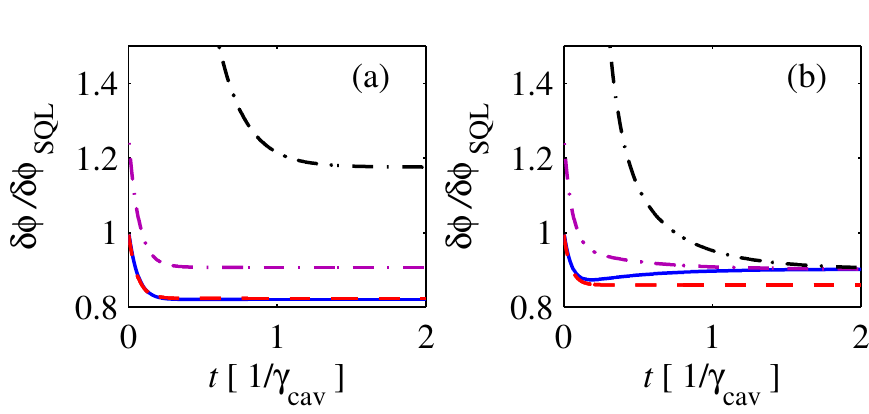}
\caption{Time evolution of the phase sensitivity for $N=10$ spins and $\Omega_-/\Omega_+=0.2$. Numerical solution of the master equation, with fully polarized (solid curves) or randomly generated (dashed-dotted curves) initial state, and mean-field approximation (dashed curves). (a) The atoms are coupled only to the cavity, i.e. $\gamma=0$. (b) Finite scattering rate $\gamma\neq 0$ with single-atom cooperativity $\chi=1$. Observe that the steady state does not depend on the initial state.}
\label{fig:dynamics}
\end{figure}

Up to this point, we have restricted ourselves to the manifold of total spin $S=N/2$. This approach is valid if the system is initially prepared in the maximally polarized state $\ket{S=N/2,S_z=N/2}$ because Hamiltonian (\ref{eq:Hint}) commutes with the total spin $\mathbf{S}^2$. Consequently, if the system is initially prepared in a state with a different total spin, the final steady state will be different and will in general contain a smaller amount of spin squeezing. This is shown in Fig. \ref{fig:dynamics}(a), where we compare the linearized solution with the exact numerical solution of the Lindblad equation $(\ref{eq:master2})$ for $N=10$ spins with different initial conditions. The linearization works well only when the spins are initially polarized, as would be expected.

So far our analysis completely disregarded the effect of spontaneous Raman scattering of the individual atoms into free space. Being a single atom process, it breaks the conservation of the total spin and thus competes with the squeezing process. To investigate the effect of spontaneous Raman scattering we introduce a spontaneous decay rate $\gamma\ll|\Delta|$ to the states $\ket{e_\pm}$. The corresponding master equation is easily obtained using standard techniques and reads
\begin{align}
 &\frac{\partial}{\partial t}\rho=-\gamma_{\rm cav}\left[\left\lbrace\mathcal{I}^\dagger\mathcal{I},\rho\right\rbrace_+-2\mathcal{I}\rho\mathcal{I}^\dagger\right] \label{eq:master3}\\
&\;-\gamma_{\rm spont}\sum_{\alpha=\sigma^\pm,\pi}\sum_{j=1}^{N}\left[\left\{{\mathcal{L}_{j,\alpha}}^\dagger\mathcal{L}_{j,\alpha},\rho\right\}_+-2\mathcal{L}_{j,\alpha}\rho{\mathcal{L}_{j,\alpha}}^\dagger\right],\nonumber
\end{align}
The first term on the RHS of (\ref{eq:master3}) is equivalent to (\ref{eq:master2}), whereas the second term describes the spontaneous Raman scattering of atom $j$ into free space with rates $\gamma_{\rm spont}=\gamma\Omega^2/\Delta^2$. The Lindblad-operators are given by $\mathcal{L}_{j,\sigma^\pm} = (\cos\theta\sigma^+ + \sin\theta\sigma^-)\sigma^\mp$ and $\mathcal{L}_{j,\pi} = \cos\theta\sigma^+ + \sin\theta\sigma^-$. The numerical solution of (\ref{eq:master3}) is shown in Fig. (\ref{fig:dynamics}b) for different initial conditions and $N=10$ spins. The spontaneous Raman scattering couples different  total spin-$S$ manifolds and, as expected, lifts the degeneracy between them. Consequently it leads to a mixed but nevertheless unique steady state, characterized by a unique density matrix, independent of the initial state, and hence presents rather a feature than a detrimental problem.

Still, Raman scattering leads to a reduction of the achievable phase sensitivity. To quantify this effect we obtain the contribution of the single-spin decay to the time evolution of the collective variables, by using the second row of (\ref{eq:master3}) and obtain
\bea
\frac{d\av{\delta S_z}}{dt} &=& -\gamma_{\rm spont} \left[\av{\delta S_z} - \sin^2(\theta) N\right],\label{eq:decay_SZ} \\
\frac{d\av{S_x^2}}{dt} &=& - 2\gamma_{\rm spont}\left(1-\half\sin(2\theta)\right) \left[\av{S_x^2} - \frac{1}4\right],\label{eq:decay_SX}
\eea
Adding (\ref{eq:decay_SZ}-\ref{eq:decay_SX}) to the contribution of the collective coupling to the cavity, eqs. (\ref{eq:cavity2_SZ}-\ref{eq:cavity2_SX}), we find that in the steady state the competition between the two processes leads to:
\bea
\av{\delta S_z}_{\infty} &=&  \frac{\sin^2(\theta)\left(\frac{N g^2}{\kappa}+ N \gamma\right)}{\frac{N g^2}{\kappa}\cos(2\theta) + \gamma}\label{eq:steadystate_SZ}\,,\\
\av{S_x^2}_{\infty} &=& \frac{1}4 \frac{\frac{N g^2}{\kappa}\left(1-\sin(2\theta)\right)+2\gamma\left(1-\half \sin(2\theta)\right)}{\frac{N g^2}{\kappa}\cos(2\theta) + 2\gamma\left(1-\half \sin(2\theta)\right)}\label{eq:steadystate_SX}\,.
\eea

\begin{figure}[t]
\centering
\includegraphics[scale=0.8]{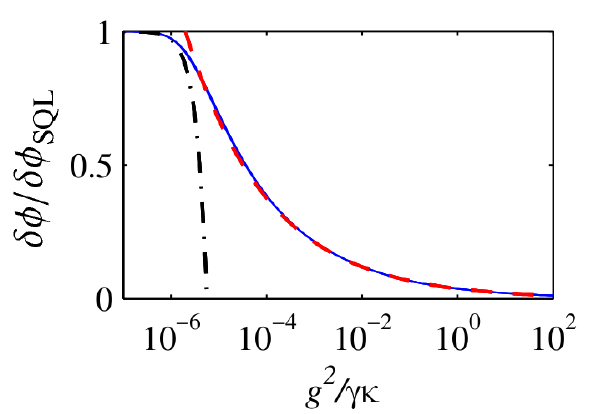}
\caption{Optimal phase sensitivity in the steady state, in the presence of single spin decay for $N=10^6$. The solid curve is given by the numerical minimization of eqs.(\ref{eq:steadystate_SZ}-\ref{eq:steadystate_SX}). The dashed-dotted and dashed curves are eqs. (\ref{eq:small}-\ref{eq:large}), respectively.}
\label{fig:analytic1M}
\end{figure}

For a given single-atom cooperativity $\chi=\gamma_{\rm cav}/\gamma_{\rm spont} = g^2/\kappa\gamma$, the optimal value of the phase sensitivity $\delta\phi$ can be obtained by numerically minimizing the corresponding expression with respect to the mixing angle $\theta$ while keeping $\chi$ fixed. The resulting curve is shown in Fig. \ref{fig:analytic1M} for $N=10^6$. To gain insight into this result, we will now separately investigate the limits of small and large single-atom cooperativity.

If single-atom cooperativity is small, but $\chi N\sim1$, the optimal spin squeezing is obtained for $\theta\ll\pi/4$. Expanding eqs. (\ref{eq:steadystate_SZ}-\ref{eq:steadystate_SX}) around $\theta\approx 0$ we obtain a phase sensitivity $\delta\phi^2 \approx \frac1N\left(1- N \chi \theta +4\theta^2\right)$ with the optimum value obtained for $\theta=N\chi/8$
\be
\delta\phi_{\rm opt} = \frac1{\sqrt{N}}\left(1-\frac{\chi^2 N^2}{32}\right)\;.
\label{eq:small}
\ee
It should be noted that due to the collective enhancement by a factor of $N$, one gains a quadratic improvement in the phase sensitivity even for small cooperativities.

If the cooperativity is larger, i.e. $\chi N\gg 1$, the optimal squeezing is obtained for $\Omega_+\approx\Omega_-$ and one may na\"{\i}vely expect the linearization to fail. However, to render the problem tractable we nevertheless perform the large-$N$ expansion and explicitly check the validity of the obtained results below. To obtain the scaling of the phase sensitivity we expand eqs. (\ref{eq:steadystate_SZ}-\ref{eq:steadystate_SX}) around $\theta  = \pi/4-\epsilon$. For small $\epsilon\sim 1/\sqrt{N}\ll 1$ and neglecting terms of order $1/N$ we find $\delta \phi^2 = \frac{1}{N}\frac{1+2N\chi\epsilon^2}{2N\chi\epsilon}$, which has a minimum at $\epsilon=\sqrt{1/2N\chi}$. This leads to a sensitivity
\be
\delta\phi =\left(\frac{2}{\chi N^3}\right)^{1/4}\;.
\label{eq:large}
\ee
{ Note that, for the optimal value of $\theta$, the ensemble is still mostly polarized, as can be checked by explicitly computing $\av{\delta S_z}\approx\frac{\chi+1}{4\chi\epsilon} \sim \sqrt{N}$. This observation provides a self-consistency check for our linearization procedure in the limit of a large ensemble, where $\sqrt{N}\ll N$. Eq. (\ref{eq:large}) shows that the achievable phase sensitivity scales as $\delta\phi\sim N^{-3/4}$ and thus offers a significant improvement over the SQL. With respect to other proposals with a similar scaling \cite{AndrePRA,Molmer,VuleticPRA}, our approach deterministically generates spin squeezing in a steady state by optical pumping and can thus potentially be more robust against external perturbations.}

A possible experimental realization could be set up in $^{87}$Rb using the clock states $\ket{F=1, m_F=0}$ and $\ket{F=2, m_F=0}$ in the $5S_{1/2}$ ground-state manifold. Circularly $\sigma_+$-polarized cavity and control fields couple these states to the states $\ket{F=1, m_F=+ 1}$ and $\ket{F=2, m_F=+ 1}$ of the $5P_{1/2}$-manifold. In this case the inclusion of the hyper-fine structure leads to an effectively larger Raman rate, as the scattering path-ways via the two hyper-fine state interfere constructively for suitably chosen detunings (see Supplementary Material \cite{Supplementary}). For current experiments \cite{Vuletic}, we have $N\approx 10^6$ and $\chi \approx 0.1$, predicting a phase sensitivity $\delta\phi/\delta\phi_{\rm SQL}\approx 0.07$, i.e. an improvement of more than one order of magnitude with respect to the SQL.

We now discuss two additional effects related to the experimental realization of the proposed model. First, in a multi-level atom (such as the suggested $^{87}$Rb), the spontaneous Raman scattering can lead to states outside of the effective four-state system. Because, as shown above, the squeezed steady state is (mixed but) unique and is achieved from any intial state, we can use additional optical pumping fields to repump the atoms into the correct configuration. In the case of $^{87}$Rb this can be achieved through linearly polarized fields resonantly driving the $F=1\rightarrow F'=1$, $F=2\rightarrow F'=2$ and $\Delta m_F=0$ transition. These fields do not couple directly to the clock states $\ket{F=1, m_F=0}$ and $\ket{F=2, m_F=0}$ and therefore act as a repumping field for the proposed four-state scheme. The repump processes can be easily incorporated in our formalism in the form of additional single-atom scattering channels \cite{Supplementary}. These processes effectively increase the single-atom scattering rate and reduce the cooperativity $\chi$. In addition, the use of a large detuning $\Delta$ makes possible to obtain repumping-rates that are much larger than the decay-rate into the external states and to minimize the loss of atoms into these states. We conclude that our results, and in particular (\ref{eq:large}), are valid even in the presence of additional repumped decay channels, provided that the cooperativity rate $\chi$ is computed with respect to the total linewidth of the excited states.

Second, in the ideal model (\ref{eq:Hint}), we assumed a spatially homogeneous coupling of the atoms to the cavity mode, which can be approximately achieved only in a ring cavity.  { In a standing-wave cavity the  off-resonant Raman coupling $g\Omega_\pm/\Delta$ depends on the position of the atoms and can be either positive or negative, leading to a cancellation of spin-squeezing. This can be solved by localizing the atoms in space and choosing an appropriate geometry for the control fields and the cavity mode, such as to make all coupling-constants having the same sign, as experimentally shown in Ref.~\cite{Vuletic}. In this case, the symmetry of the collective state is preserved, leading only to a small reduction of the effective spin cooperativity.}

To summarize, we presented a method for the deterministic generation of spin squeezed states in a driven ensemble of effective two-state atoms in a strongly dissipative cavity. The generating process can be understood as optical pumping into a non-equilibrium steady state of the atom that at the same time is a dark-state with respect to the cavity decay. Introducing spontaneous Raman scattering we showed that the squeezed steady state is unique and does not depend on the initial state of the system. We discussed the effect of the single atom spontaneous Raman scattering on the achievable phase-sensitivity and found that it scales favorably with the single-atom cooperativity, indicating that the present method can be of direct importance for, e.g., optical atomic clocks.

We thank  S. Bennett, M. Fleischhauer, M. Hafezi, N. Yao, and P. Zoller for useful discussions and NSF, CUA, DARPA, ARO MURI and Packard for financial support. J.O. acknowledges support by the Harvard Quantum Optics Center. The numerical solution of the master equations were performed on the Odyssey cluster supported by the FAS Science Division Research Computing Group at Harvard University.

\end{document}


\title{Dissipative Preparation of Spin Squeezed Atomic Ensembles in a Steady State -- Supplementary Material}

\author{Emanuele G. Dalla Torre$^{1}$}
\thanks{E.G. Dalla Torre and J. Otterbach equally contributed to this work}

\author{Johannes Otterbach$^{1}$}
\thanks{E.G. Dalla Torre and J. Otterbach equally contributed to this work}

\author{Eugene Demler$^{1}$,Vladan Vuletic$^{2}$, Mikhail D. Lukin$^{1}$}

\affiliation{$^{1}$Physics Department, Harvard University, Cambridge 02138, MA, USA, $^{2}$Physics Department, Massachussetts Institute of Technology, Cambridge 02139, MA, USA}

\date{\today}

\maketitle


\section{Experimental implementations and optical repumping}

\begin{figure}[t]
 \centering
 \epsfig{file=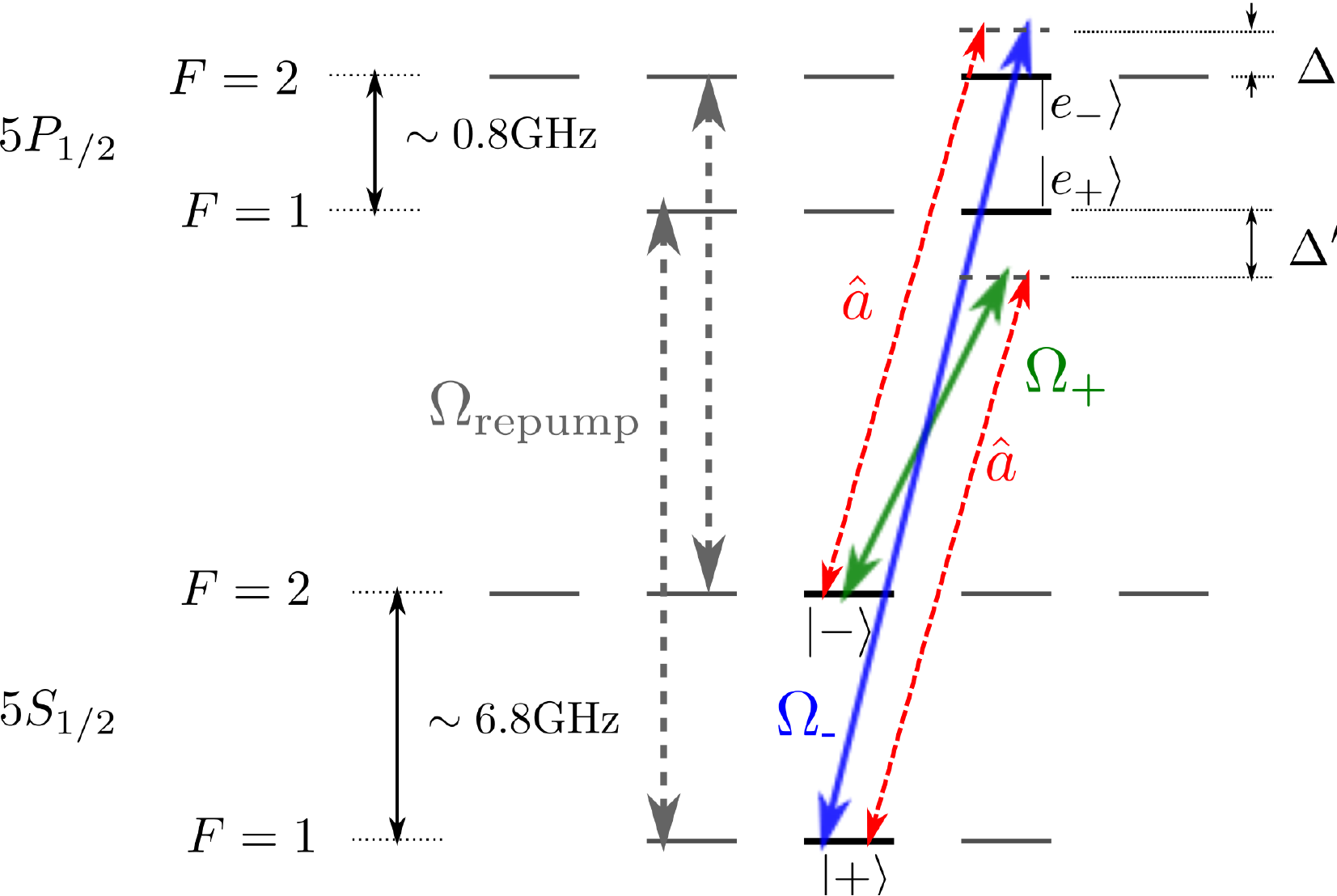,width=.45\textwidth}
 \caption{Experimental level-scheme in $^{87}$Rb using the clock states.}
 \label{fig:Rb-linkage}
\end{figure}

In this section we investigate in detail the proposed experimental realization. Fig. \ref{fig:Rb-linkage} shows the relevant level-scheme of the $^{87}$Rb $5S_{1/2}$ and $5P_{1/2}$ manifolds \cite{Steck-Data}. The cavity mode is a circularly polarized mode, which can be achieved using a running wave cavity with an optical element breaking the symmetry between the two degenerate polarization. Alternatively one can use a non-polarized cavity mode and apply a magentic field that lifts the Zeeman degeneracy. In this way, the relevant scheme can be made two-photon resonant, while keeping the complementary one (i.e. with opposite polarizations) two-photon off-resonant and thus strongly suppressed. If the control fields propagate along the cavity axis, the free spectral range of the cavity has to be choosen such that the cavity-mode and the control-fields are supported, or the power of the control-fields has to be made sufficiently large.

To derive the effective Hamiltonian (1) of the main text, we employ the Heisenberg-Langevin formalism \cite{Louisell}. Working far off single-photon resonance, i.e. $|\Delta|,|\Delta'|\gg \gamma$, we can neglect the spontaneous decay and consequently also the Langevin-noise forces. For simplicity of the calculation we assume that we can realize two independent $\Lambda$-schemes according to Fig. (\ref{fig:Rb-linkage}) and can neglect the respective second state. While this approximation is only true if the detunings from the respective excited state is smaller than the hyperfine-splitting in the excited state manifold, Hamiltonian (1) is valid for more general conditions, when Raman processes mediated by multiple virtual states, can be added up \cite{Harris1997, Harris1999}. The validity of this approximation will be further discussed below. With these approximations the equations of motion of the atomic variables are given by
\begin{align}
 \pdt\S{+e_+} =& -i\Delta' \S{+e_+}+i\Omega_- \S{+-}+ig\S{++}\a \label{eq:eom_spin_0}\\
 \pdt\S{+e_-} =& i\Delta \S{+e_-}+i\Omega_+ \S{++}+ig\S{+-}\a \\
 \pdt\S{-e_+} =& -i\Delta' \S{-e_+}+i\Omega_- \S{--}+ig\S{-+}\a \\
 \pdt\S{-e_-} =& i\Delta \S{-e_-}+i\Omega_+ \S{-+}+ig\S{--}\a \\
 \pdt\S{+-} = & -i\Omega_+\S{e_- -}+i\Omega_-\S{+\e+} \nonumber \\&-ig\S{e_+ -}\a+ig\ad \S{+e_-} \label{eq:eom_spin_1}
\end{align}
Adiabatic elimination of all variables involving the excited states and subsequent insertion into Eq. (\ref{eq:eom_spin_1}) yields
\begin{align}
\pdt\S{+-}= &\; i\left(\frac{\Omega_+^2}{\Delta}+\frac{\Omega_-^2}{\Delta'}\right)\S{+-}\\
&+ig\left(\frac{\Omega_+}{\Delta}\a -\frac{\Omega_-}{\Delta'}\ad \right)\S{z} -i\left(\frac{g^2}{\Delta}+\frac{g^2}{\Delta'}\right) \ad\a\S{z} \nonumber
\end{align}
We see that the off-resonant Raman coupling leads to an additional Stark-shift $\sim\left(\frac{\Omega_+^2}{\Delta}+\frac{\Omega_-^2}{\Delta'}\right)$ which, however, can be compensated by engineering a properly chosen two-photon detuning. The shift of the cavity resonance $\sim\left(\frac{g^2}{\Delta}+\frac{g^2}{\Delta'}\right)$ can be neglected as in the steady state the photon number in the cavity is zero. Neglecting these two frequency shifts, the above equation is equivalent to the effective single particle Hamiltonian
\begin{align}
 h_{\rm eff}=-ig\left(\frac{\Omega_+}{\Delta}\a -\frac{\Omega_-}{\Delta'}\ad \right)\S{-+}+\text{h.c.}
\end{align}
Defining the collective operators $\mS_+ = \sum_j \S{+-}^{(j)}$, where the sum runs over all atoms, we find the final result
\begin{align}
 H_{\rm eff} = \sum_j h^{(j)}_{\rm eff} = \frac{g}{\Delta}\ad \left(\Omega_-\frac{\Delta}{\Delta'}\mS_--\Omega_+\mS_+ \right)+\text{h.c.}
\end{align}
Redefining $\Omega_-\frac{\Delta}{\Delta}'\rightarrow -\Omega_-$ results in Hamiltonian (1) of the main text. Hence, by adjusting relative phase and strength of the control fields the proposed Hamiltonian can be engineered.

In the derivation of Eqs. (\ref{eq:eom_spin_0}-\ref{eq:eom_spin_1}) we assumed the two Raman transitions involving $\Omega_+$ and $\Omega_-$ to be independent. This approximation is valid only if both the detunings $\Delta$ and $\Delta'$ are smaller than the hyperfine-splitting $\delta_{\rm HFS}$ in the excited state manifold. However, for the proposed realization in $^{87}$Rb, the above condition cannot be fulfilled since the frequency of the cavity fixes the difference of the detunings to be equal to the difference of the hyperfine-splittings in the ground and excited state. Hnece we can at most choose one of the detunings to be smaller than $\delta_{\rm HFS}$. Nevertheless, as we will show now, in the system under consideration the coupling to the additional state interferes constructively giving rise to an effectively larger Raman scattering rate and thus rendering valid our approach.

We first consider the Raman transition involving $\Omega_-$ and denote the cavity dipole matrix-elements associated with the transitions $\ket{-}\rightarrow\ket{e_-}$ and $\ket{-}\rightarrow\ket{e_+}$, respectively, as $g$ and $g_2$. Analogously, we define the Rabi frequency of the control field driving the transitions $\ket{+}\rightarrow\ket{e_-}$ and $\ket{+}\rightarrow\ket{e_+}$, respectively, as $\Omega_-$ and $\Omega_{-,2}$. Generalizing the above approach we find that the total Raman scattering rate is then given by
\begin{align}
 \Omega_{-,\rm tot} = \frac{g\Omega_-}{\Delta}+\frac{g_2\Omega_{-,2}}{\Delta+\delta_{\rm HFS}}\;.\label{eq:omegatot}
\end{align}
The coupling matrix-elements can be expressed as $g=\wp_{-,e_-}\sqrt{\omega_c/2\hbar\epsilon_0 V}$ and $g_2=\wp_{-,e_+}\sqrt{\omega_a/2\hbar\epsilon_0 V}$ where $\wp_{-,e_\pm}$ is the dipole matrix-element of the respective transition. For the $^{87}$Rb D1 line these matrix-elements are given by $\wp_{-,e_-} =\sqrt{1/4}\langle J=1/2||er||J'=1/2\rangle $, $\wp_{-,e_+} =\sqrt{1/12} \langle J=1/2||er||J'=1/2\rangle $, with $\langle J=1/2||er||J'=1/2\rangle$ being the reduced matrix-element  \cite{Steck-Data}. Analogously, the Rabi frequencies of the control fields are given by $\Omega_- = \wp_{+,e_-} E_-/\hbar$ and $\Omega_{-,2} = \wp_{+,e_+} E_-/\hbar$, where $E_-$ is the electric field amplitude of the control field. The matrix-elements are given by $\wp_{+,e_-} =-\sqrt{1/4}\langle J=1/2||er||J'=1/2\rangle $, $\wp_{+,e_+} =-\sqrt{1/12} \langle J=1/2||er||J'=1/2\rangle$. Plugging these results into Eq.~(\ref{eq:omegatot}) we find
\begin{align}
 \Omega_{-,\rm tot} = &\left(\frac{1}{4\Delta}+\frac{1}{12(\Delta+\delta_{\rm HFS})}\right) \times \nonumber \\
 &\;\frac{E_-}{\hbar}\sqrt{\frac{\omega_c}{2\hbar\epsilon_0 V}}\langle J=1/2||er||J'=1/2\rangle
\end{align}
Thus as long as $\Delta$ and $\Delta+\delta_{\rm HFS}$ are of equal sign the two Raman rates constructively interfere giving rise to an effectively larger total Raman scattering rate. An analogous calculation for the transition involving the second control field $\Omega_+$ yields a similar result. This shows, that as long as we choose the detunings such that we are always outside the hyper-fine structure, i.e. either blue or red detuned with respect to all hyperfine-states, we can always find a Raman scheme that gives rise to the effective Hamiltonian (1) of the main text.

\begin{figure}[t]
\centering
\epsfig{file=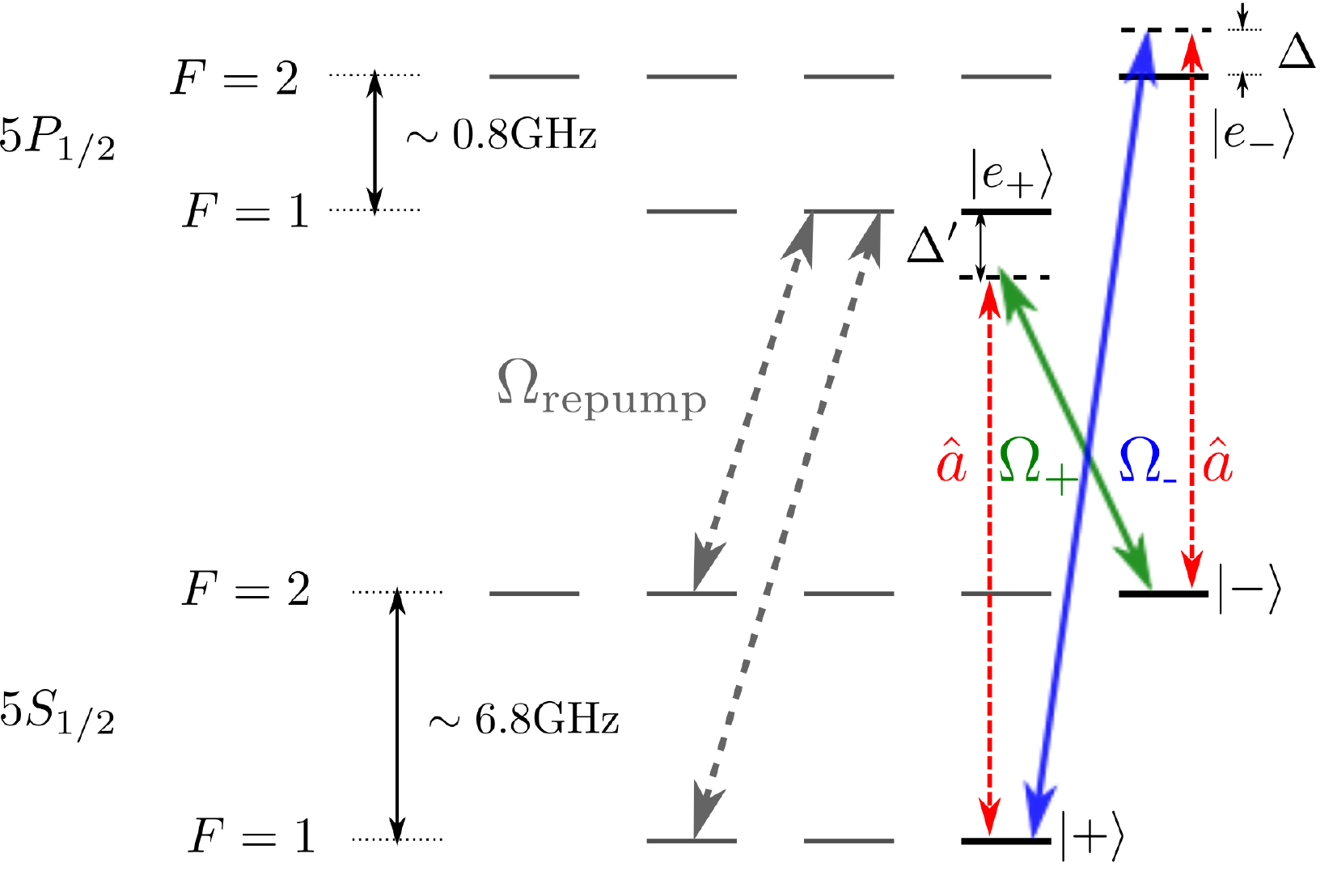,width=.45\textwidth}
\caption{Alternate experimental level-scheme using the stretched states of $^{87}$Rb.}
\label{fig:Rb-linkage-2}
\end{figure}

As an alternate coupling scheme one could use the strechted states $\ket{F=1, m_F=1}$ and $\ket{F=2, m_F=2}$ in the $5S_{1/2}$ ground-state manifold as implementation of the spin-system (cf. Fig.\ref{fig:Rb-linkage-2}). Using linear polarized cavity fields along with circular polarization for the control fields results in the same effective Hamiltonian, once relative phase and strength of the control fields are chosen. However, this setup needs a different experimental geometry with the pump-beams entering the cavity from the side and the atoms trapped in a $\lambda$-lattice, to ensure that the spatial variations of the coupling matrix-elements have the same sign for all atoms.

Let us now address the question of spontaneous decay of the atoms into state outside the effective spin-system. This leads to two effects (i) a renormalization of the single atom cooperativity and (ii) to a slow decay of the number of atoms contributing to the spin-system. To see the first effect we note, that the presence of additional decay channels gives rise to a larger bandwidth of the excited states $\ket{e_\pm}$. Hence we have to replace the original decay rate $\gamma$ by a decay-rate that also incorporates the decay rate $\gamma_0$ into other states, i.e. $\gamma\rightarrow\gamma+\gamma_0$, which, in turn, renormalizes the single-atom cooperativity to smaller values. To circumvent the second effect, we propose to optically pump the states outside the spin-system back into the spin-states. For the realization involving a clock transition (cf. Fig \ref{fig:Rb-linkage}), we note that the transition between the two clock-states $\ket\pm = \ket{F=1,2,m_F=0}$ to the excited states $\ket{F'=1,2,m_F=0}$ is forbidden. Hence any resonant linearly polarized light will bring the population back into the two clock states in analogy to optical pumping with linearly polarized light. Similar schemes can also be found for an experimental implementation using the strechted states of $^{87}$Rb.
The population in the external states is then given by the ratio between the decay-rates leading out of the spin-system and the repumping rate. In the limit of weak driving, i.e. $\Omega_{\rm repump}\lesssim \gamma_0$, where $\Omega_{\rm repump}$ is the Rabi frequency of the repump, and $\Omega_+\lesssim |\Delta'|$, $\Omega_-\lesssim |\Delta|$, these rates are given by $\Gamma_{\rm out,\pm}=\gamma_0 (\Omega_\pm/\Delta_\pm)^2$ and $\Gamma_{\rm repump}= \Omega_{\rm repump}^2/(\gamma_++\gamma_-)$. For large detuning, i.e. $\Delta^2\gg\gamma_0(\gamma_++\gamma_-)$, this ratio can be made arbitrarily small, thus leaving only a small population in the external state.

\bibliographystyle{plain}
